# Arabic Interface Analysis Based on Cultural Markers

Mohammadi Akheela Khanum[1], Shameem Fatima[2], Mousmi A.Chaurasia[3]
Information Technology Department, King Saud University
Riyadh, Kingdom of Saudi Arabia
Email:[1]kakheela@ksu.edu.sa, [2]sfatima@ksu.edu.sa, [3]mchaurasia@ksu.edu.sa

**Abstract**
This study examines the Arabic interface design elements that are largely influenced by the cultural values. Cultural markers are examined in websites from educational, business, and media. Cultural values analysis is based on Geert Hofstede's cultural dimensions. The findings show that there are cultural markers which are largely influenced by the culture and that the Hofstede's score for Arab countries is partially supported by the website design components examined in this study. Moderate support was also found for the long term orientation, for which Hoftsede has no score.

## 1. Introduction

Human-Computer Interaction (HCI) is both an **art and a science**. The interdependence of a software system's functionality and its interface means that software designers cannot afford to favor one over the other. If the interface is well designed, it will allow the system's functionality to support the user's task. However, if the interface is inadequate, the functionality is obscured and users will have trouble accomplishing their task [1].With the growth and expansion of the Internet and World Wide Web, the number of non-English speaking users of Internet also is growing rapidly. For instance, the number of Arabic and Chinese speaking users has grown 1907.9% and 1087.7% respectively [2]. These figures are a clear indication that web connects various regions and various people across the world. The "one size fits all" formula no more holds for the web interface design. Therefore, the cultural context of user interface design has become an important issue to be considered while developing interfaces for different cultures. In this paper we discuss issues regarding the influence of culture on Arabic websites design. Arabic websites from three Arab countries (Saudi Arabia, UAE, and Kuwait) are taken for our analysis. We consider different educational websites, media websites and business websites, and analyze them according to Geert Hofstede's theory of culture dimensions [3]. The results of this study can be used by the web designers to design more localized interfaces for the Arab users.

## 2. Cultural Issues in Interface Design

### 2.1 Hofstede's cultural Values

There are many models of culture that are used by the researchers and practitioners which can help in studying and designing websites across cultures. Geert Hofstede's model is one of the most widely used models for studying the cross cultural challenges in the design of interfaces. Hofstede cultural dimensions are based on a large sample of employees from 40 countries from the large multinational IBM, whom he studied from 1960's, 70's and 80's. Hofstede's model comprises of five cultural dimensions. These dimensions comprise: *Power Distance*-the extent to which the less powerful members of organizations expect and accept that power is distributed unequally; *Individualism*-the extent to which individuals are integrated into groups; *Masculinity*-assertiveness and competitiveness versus modesty and caring; and *Uncertainty Avoidance*-intolerance for uncertainty and ambiguity. A fifth dimension, *Long-Term Orientation*-the degree of future orientation was added later on in 1982 when Hofstede expanded his model to include 10 more individual countries and three regions. As a group, Arab countries scored high on Power Distance (80), Uncertainty Avoidance (68), and Masculinity (52) dimensions, while scoring low on the Individualism (38) dimension. The only dimension that does not have any scores for these countries is the Long-/Short-Term Orientation dimension [3].

### 2.2 Related studies in non-Arab cultures

Researches in the past have focused on the effect of culture on the interface usability and influence of culture on the interface design. Some of the literature reviewed by us is as follows:
Radmila et al [4] in their study on developing UK and Korean cultural markers pointed to the general issues of the cross-cultural web design. To do this, they defined a checklist of relevant design elements which are supposed to be culturally specific design elements, called as the cultural markers [5]. The cultural markers they used included verbal attributes such as language and format (time, date, addresses, currency, printing format and size, units of measurements), visual attributes which comprised of images, color, text, layout, and audiovisual attributes which includes sound, animation and 3D. The results indicate that some of the cultural markers were having same values across the UK and Korean websites and hence the authors concluded that such elements are not candidates for cultural markers and they may not influence the webpage design and usability.

Authors in [6] conducted a study on comparison of Malaysia and Britain local cultural values through the website analysis. They applied Hofstede's individualism/collectivism and power distance, and Hall's high/low-context cultural dimensions, to the various websites which included the university websites, tourism websites and bank websites. The result indicates that the cultural values presented in the local websites of the two varying cultures match the research of both Hofstede and Hall regarding the cultural differences between the countries.

Tong & Robertson[7] in their research on the political and cultural representation in Malaysian websites, adopted power distance from Hofstede's model of cultural analysis and Aaron Marcus's approach to multi-dimensional web-interface analysis to identify current representation of multicultural Malaysia. They used cultural marker's model (CMM) to investigate cultural inclusion. The results suggest that, it is not easy for designers to develop a sophisticated understanding of culturally sensitive visual interface design. Although there are many existing frameworks and theories, it is difficult for designers to identify the appropriate model for a particular multicultural society.

Kim & Kuljis [8] conducted a study on identification of elements that can be attributed to culture in the website design. They compared the South Korean and UK's charity websites based on Greet's Hofstede's theory of culture dimensions. The results show that there are some differences and preferences in the websites design that are mostly related to whether the websites employ multimedia and provide facilities for user input.

2.3 What do we know about Arab cultural markers?

Very fewer studies in the past have focused on the Arab interface design issues. Some of them are as follows:

Khusman et al. [9] proposed a model that includes cultural variables, which largely influence the user's acceptance behavior for the e-business websites in Arab countries. A field study was conducted in three main tourist sites of Jordan. Tourists from UK and Arab countries were used as the target group for this study. The questionnaire consists of questions related to the respondent's background and possible factors that may influence their acceptance and usage of e-business websites. The results suggests that e-business websites developed for low power distance, low uncertainty avoidance, high individualism and high masculinity cultures (like the Western cultures) are not optimally suited for Arab cultures which involves high power distance, high collectivism, low masculinity and high uncertainty avoidance. The analysis contradicts the results of Hofstede, which suggest that Arab cultures display a lower masculinity than western cultures.

Research by Aaron Marcus & Associates [10] in the year 2009, discusses issues regarding the influence of culture on Arabic websites. They analyzed three Arab countries, the Jordan, Egypt and the United Arab Emirates. The analysis was on the educational websites based on Geert Hofstede's theory of culture dimensions and Marcus (author) theory of user interface components. Marcus components include the metaphors, the mental model, navigation, interaction, and presentation styles of interface design. The results points out that Arabic websites need to consider some changes, such as, to add more representative pictures, more multimedia components, more links to the external websites, and more multilingual contents.

Study conducted by Khashman & Large [3] examine the design characteristics of government web interfaces from three Arab countries using Hofstede's cultural dimensions. Organizational and graphical elements from 30 ministry websites from Egypt, Lebanon and Saudi Arabia were examined using content analysis. Element frequency scores were correlated with Hofstede's dimensions and interpreted based mainly on the model developed by Marcus and Gould. The results suggest that Hofstede's model of culture does not fully reflect the design characteristics of Arabic interfaces.

## 3. Method

In order to understand the user interface requirements for any population, it is important to analyze the cultural influence on the acceptance of technologies. The primary aim of this research is therefore is to explore the cultural values of the Arab countries through analyzing their website designs. We used systematic analysis of cultural markers to find out how much the websites design conforms to the Hofstede's cultural dimensions. Also, we analyzed the cultural markers that are most prominent in Arab culture. We applied the list on a selection of 27 web pages from three countries in the Arabian Gulf region. These include Saudi Arabia, United Arab Emirates and Kuwait which are believed to have similar cultures. Most of the previous studies have focused on only one genre, either educational, Government or the tourist websites. Our perception is that the more the number of genres used, clearer would be the insight. Therefore, we chose Web pages from the following three genres (9 Web pages per

genre)

- Education
- News and Media
- Business

These genres are chosen on the basis that the websites for these genres are created by and for the people belonging to the local culture and consequently reflecting the socio-cultural and technological characteristics of their culture so as to be successful in providing the services to the target population. We selected the home pages for analysis, as they may contain many central elements of web design.

Table 1: List of Cultural Markers Used

| | | |
|---|---|---|
| **COLOR** | Background | |
| | Text | |
| | Link | Visited Link |
| | | Unvisited ink |
| | Menu Background | |
| | Menu Font | |
| | Picture | |
| | Logo | |
| **LAYOUT** | Menu Click | |
| | Menu Place | |
| | Scroll Bar | |
| | Page Orientation | |
| | Graphic / Image | |
| | Logo | |
| **TEXT** | Typeface | Title |
| | | Body |
| | | Menu |
| | Text Size | Title |
| | | Body |
| | | Menu |
| **LANGUAGE** | Native (*web site in native Arabic language only*) | |
| | Foreign (*non-Arabic interface only*) | |
| | Multiple (*bilingual Arabic/English or additional languages*) | |
| **# OF LINKS** | Internal | |
| | External | |
| | Total | |

## 3.1 Cultural Markers

The list of cultural markers that we used is listed in Table 1. Cultural markers were based on the components proposed by Barber and Badre [5]. Studies by Galdo [11], Fernandes [12] and Russo and Boor [13], shows that cultural factors such as the icons, color, symbols and language are essential elements to be considered while designing website. The web pages were analyzed based on five characteristic elements of web design which include Layout, Color, Text, Language, and Number of Links. These cultural markers have been used in several HCI studies which examine cultural usability [4, 5].

## 3.2 Stimuli

Variability in the sample being examined was considered to ensure that web pages were representative of the culture in the Arabic Gulf Region.

### 3.2.1 Education

Web pages for Education websites included nine organizations of higher education. These nine university web pages are categorized by Country and described in the figure below. For Saudi Arabia, web sites for three universities were selected. The first is the King Saud University, the first and the largest university in the Kingdom of Saudi Arabia. The second is King Fahad University of Petroleum and Mineral Sciences. Third is the King Abdulaziz University in the city of Jeddah in the Western province of the Kingdom. Fig. 1 shows the screenshots of the three education websites from Saudi Arabia

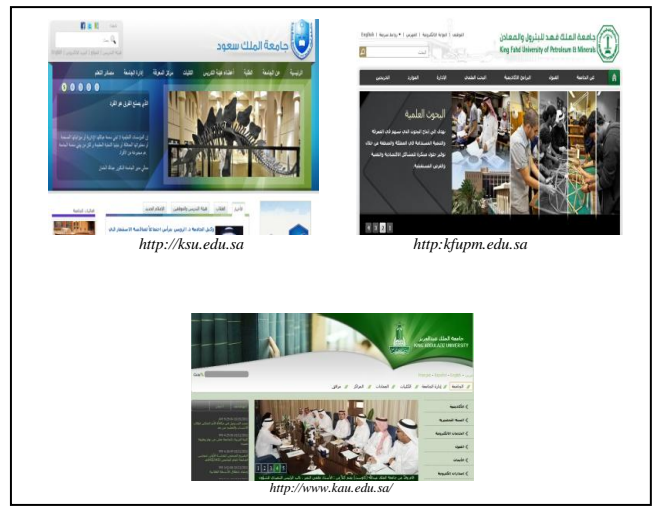

*http://ksu.edu.sa*   *http:kfupm.edu.sa*

*http://www.kau.edu.sa/*

Fig. 1 Education Websites from Saudi Arabia

For the UAE websites, three universities were selected covering both private and public higher educations. They include the UAE University, the leading and pioneering educational institution in the region, the Zayed University in Dubai and Abu Dhabi and the Sharjah University, a leading institution for higher learning located in Sharjah. Fig.2 contains the screenshots of the education websites from UAE.

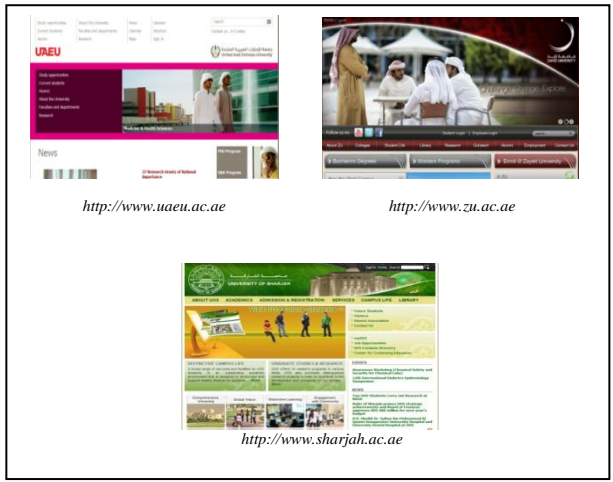

*http://www.uaeu.ac.ae*     *http://www.zu.ac.ae*

*http://www.sharjah.ac.ae*

Fig. 2 Education websites from UAE

For Kuwait, the three university websites comprising of the Kuwait University, the Global University of Science and Technology, and Kuwait Institute for medical specialization, are analysed for the cultural markers. Fig.3 depicts the homepages of the three education websites from Kuwait.

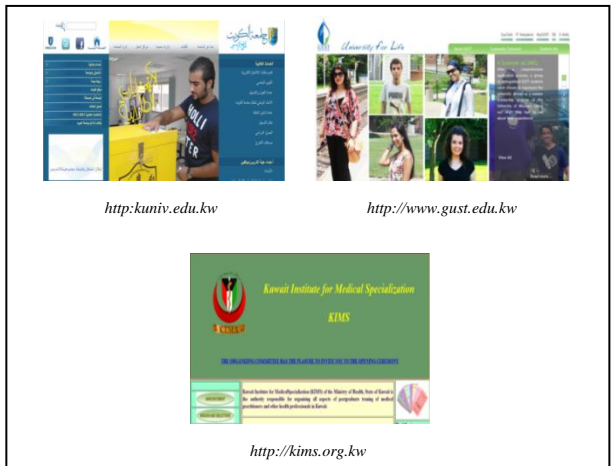

*http:kuniv.edu.kw*     *http://www.gust.edu.kw*

*http://kims.org.kw*

Fig. 3 Education websites from Kuwait

### 3.2.2 News and Media

From Saudi Arabia, three prominent and widely read newspaper's websites were analyzed. The Aljazirah and the Okaz Arabic newspaper were established in 1960, whereas the Asharq Alawsat was launched in London in 1978. Aljazirah is the first publication in Saudi Arabia. Okaz is the Arabic sister newspaper of Saudi Gazette. Asharq Alawsat, the first Arabic daily newspaper to execute satellite transmission for simultaneous printing in on four continents in 12 cities. Fig.4 displays the screenshots of the three media websites from Saudi Arabia.

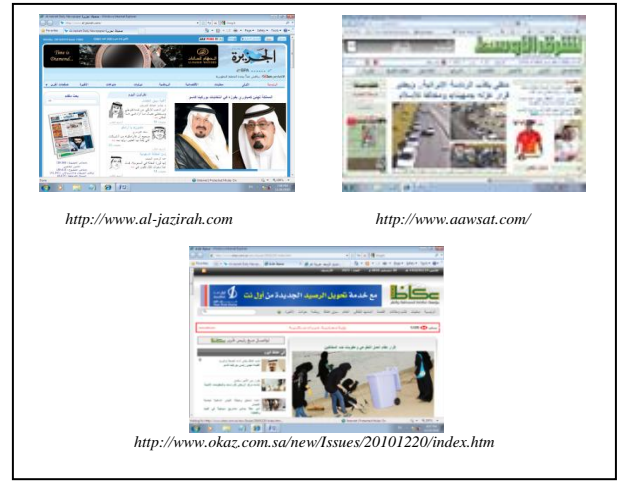

*http://www.al-jazirah.com*     *http://www.aawsat.com/*

*http://www.okaz.com.sa/new/Issues/20101220/index.htm*

Fig. 4 Media websites from Saudi Arabia

From UAE, we considered three mostly read newspapers. The Alittihad is the oldest newspaper in UAE. Alkhaleej is a daily Arabic language newspaper published in Sharjah and established in 1970.Albayan was established in 1980 by the government of Dubai. The three media websites of UAE are displayed in Fig.5.

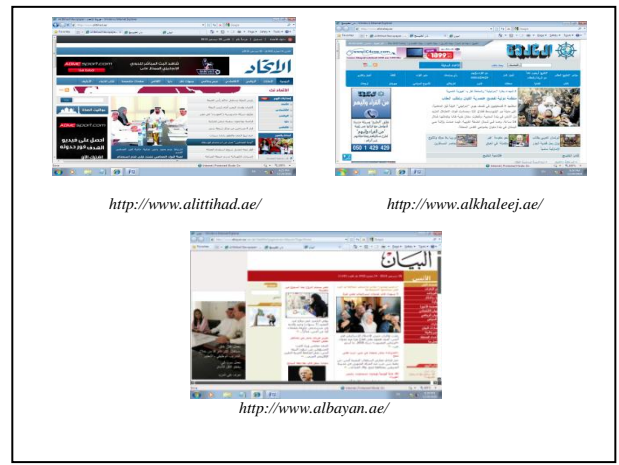

*http://www.alittihad.ae/*     *http://www.alkhaleej.ae/*

*http://www.albayan.ae/*

Fig. 5 Media websites from UAE

Three mostly read newspapers from Kuwait are considered for the analysis. They include the Alanba, Alqabas, and the Alrai. Fig.6 is the screenshots of three media websites from Kuwait.

### 3.2.3 Business

Nine Business organizations' websites where chosen for analysis. From Saudi Arabia three websites which include Saudi Aramco one of the leading oil and gas company, Saudi Basic Industries Corporation, leading manufacturers of chemical, fertilizer, Plastics and metal and Arab Petroleum Investments Corporation. Websites screenshots of three business organizations is shown in Fig.7.

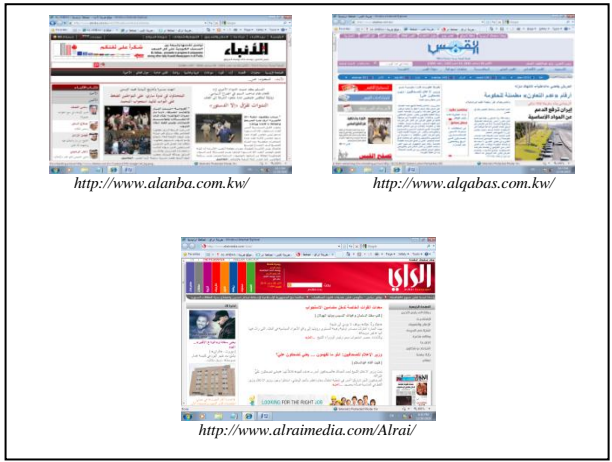

Fig. 6 Media websites from Kuwait

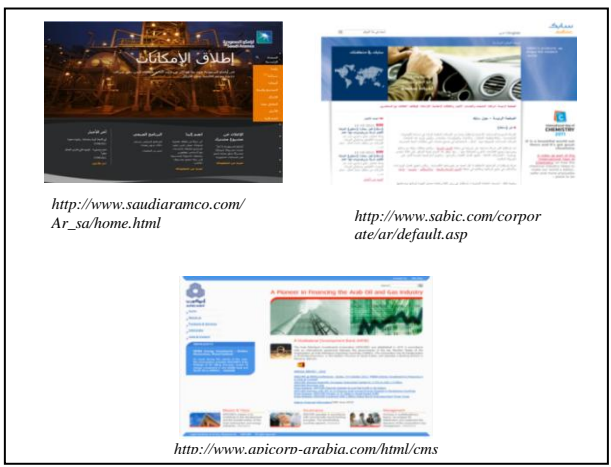

Fig. 7 Business websites from Saudi Arabia

Three websites from UAE includes the Abu Dhabi National Oil Company, which is a petrochemical company, the Crescent Petroleum and the Abu Dhabi Company for Onshore Oil operations. Fig.8 shows the screenshots of three business websites from UAE.

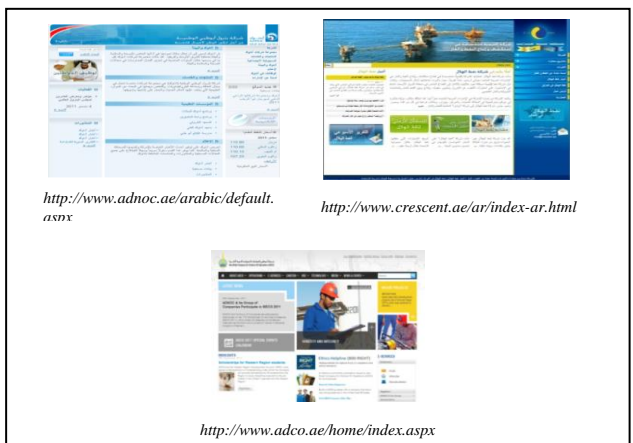

Fig. 8 Business websites from UAE

From Kuwait we have selected the Kuwait National Petroleum company, the Kuwait petroleum Corporation and the Kuwait oil company. Fig.9 is the display of the business websites from Kuwait.

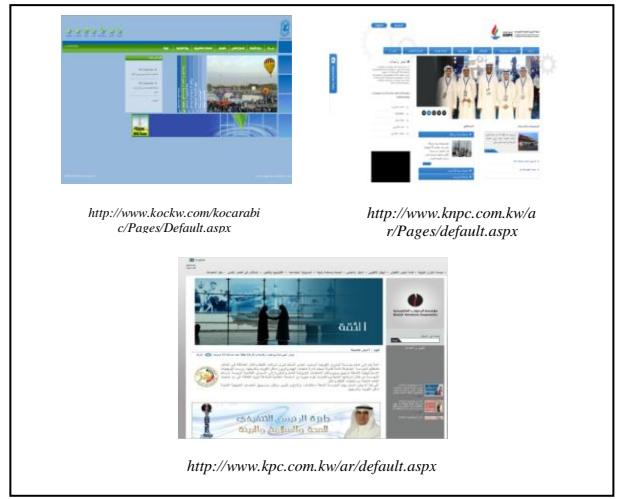

Fig. 9 Business websites from Kuwait

## 4. Result

The analysis was conducted on these websites and 3 raters were used to examine the cultural markers.

### 4.1 Markers for color

Color is one of the strongest cultural markers. Color in our analysis refers to the colors of webpage background, typography, links pictures and logos. Different colors represent different meanings in different cultures, for example, Red color in China means happiness but danger in the US [5]. Therefore, use of color in interface design may have a greater impact on the user's satisfaction and expectations [7]. The cultural markers analysis of the result shows that white color, which symbolizes purity and peace in the Arabian culture, has been used prominently as background color (85%) as well as the menu font color (48%). The color blue, which has been suggested to indicate protection in Arabian culture [15], has been used as the visited and unvisited links color (55%) and also as the menu background color (26%) in most of the web pages that we analyzed. Logo, which represents the historical background of the country, can be found on each web site. Multiple colors are usually used for the logo with green and blue dominating. The color markers results summary is depicted in Table 2.

Table 2: Result summary for Color Markers

| Color | | | | | | | |
|---|---|---|---|---|---|---|---|
| **Background** | **Text** | **Link** | | **Menu** | **Menu font** | **Image** | **Logo** |
| | | *Visited* | *Unvisited* | | | | |
| White=23 | Gray=14 | Blue=14 | Blue=15 | Blue=7 | White=13 | Multiple=27 | Multiple=13 |
| Maroon=1 | Black=6 | Green=4 | Green=4 | White=5 | Blue=6 | | Blue=5 |
| Green=1 | White=2 | Gray=3 | Gray=3 | Gray=4 | Gray=3 | | Green=5 |
| Blue=2 | Green=2 | Purple=1 | Red=2 | Green=3 | Green=2 | | Black=3 |
| | Multiple=2 | Red=1 | Maroon=1 | Black=2 | Black=2 | | White=1 |
| | Blue=1 | Maroon=1 | Orange=1 | Orange=1 | Red=1 | | |
| | | Orange=1 | Multiple=1 | Maroon=1 | | | |
| | | Multiple=1 | | Red=1 | | | |
| | | | | Yellow=1 | | | |
| | | | | Multiple=1 | | | |

## 4.2 Markers for layout

According to Yu and Roh [14]," appropriate design layout provides web visitors with a contextual and structural model for understanding and accessing information". Layout of the web pages which includes the menu placement, the place of the scroll bar, the page orientation and the place of the logo, identifies the cultural preferences. All the surveyed websites are vertically oriented with a side scroll bar. Static menus prominently placed on the top can be found on most (52%) of the websites. Some websites also have menu placed both on the top as well as on the left and right side of the page. Images on the websites are placed prominently on the top; some images are also found on different parts of the webpage. Images show the official logos, official buildings, students, and official authorities (deans, chairmen, and the founder). The result of layout analysis is shown in Table 3.

Table 3: Result summary for Layout Markers

| Layout | | | | | |
|---|---|---|---|---|---|
| **Menu click** | **Menu place** | **Scroll bar** | **Page orientation** | **Graphic/Image** | **Logo** |
| Static=14 | Top horizontal=18 | Vertical=27 | Vertical=27 | Top=10 | Top Right=21 |
| Drop Down when mouse hovers= 9 | Right vertical=4 | | | Everywhere=9 | Top Left=5 |
| Drop Down = 3 | Center=2 | | | Center=1 | Center=1 |
| | Multiple=2 | | | left side=5 | |
| | Vertical Left=1 | | | Right side=1 | |
| | | | | No graphic/image=1 | |

## 4.3 Markers for text

Text is another important cultural marker which strongly represents the cultural preferences. The type of fonts used in the web pages has an impact on its usability. In our survey of the websites, we found the font types Tahoma and Arial are preferred fonts for the title, body and menu. The text size for the body, menu and title ranges from 12px to 15px.The text analysis results are depicted in Table 4.

Table 4: Result summary for Text Markers

| Text | | | | | |
|---|---|---|---|---|---|
| Type face | | | Text size | | |
| *Text* | *Body* | *Menu* | *Text* | *Body* | *Menu* |
| Image=19 | Tahoma=7 | Image=7 | Image=19 | 11px=3 | Image=7 |
| Normal=5 | Arial=5 | Arial=4 | Medium=4 | 12px=7 | 10px=2 |
| Tahoma Bold=1 | Multiple=4 | Multiple=4 | 32px=1 | 13px=3 | 11px=1 |
| Italic Bold=1 | Arabic Transparent=4 | Tahoma=3 | 15px=2 | 14px=2 | 12px=2 |
| Arial=1 | Simplified Arabic=2 | Normal=3 | 12px=1 | 15px=4 | 13px=4 |
| | Normal=2 | Transparent Arabic=2 | | 16px=3 | 14px=3 |
| | Traditional Arabic=1 | Bold=1 | | 18px=1 | 15px=3 |
| | Image=1 | Myriad Pro=1 | | 19px=1 | 16px=2 |
| | Verdana=1 | Times New Roman=1 | | 21px=1 | 20px=1 |
| | | Simplified Arabic=1 | | Image=1 | 26px=1 |
| | | | | | 29px=1 |

## 4.4 Markers for languages and links

The most distinctive cultural symbol is language. In everyday usage, Arabic is most commonly shared language for Arab society although people have different local dialects and culture, the learning of English and its use is common in everyday life. The bilingual websites are developed probably due to the fact that there are minorities and many foreign workers in these countries who do not speak Arabic. The language preferences have been tested based on native (native Arabic language only), foreign (non-Arabic interface only) and multiple (bilingual Arabic/English or additional languages). The result analysis shows that most of the websites are bilingual (52%), 30% of the websites have only Arabic interfaces with a majority being the media websites. Very few (18%) have only English interface. Apart from this, a fewer websites also have interfaces in other languages such as Spanish and French. The websites having only English interface could pose a barrier for the Arabic speakers. Links are used everywhere, in navigation, in banners, and in graphics. We evaluated the total number of links (sum of internal and external links) for each of the website using the link counter available on [16]. We considered internal links as those links which will open in the same window and the external links will open in a new window. 93% of the total links open in the same browser window and only 7% of the links open in the new browser window. These figures are indicative of the sequential way of Arabic culture preference in solving the problems. The result summary for languages and links can be found in Table 5.

Table 5: Result summary for Languages and Links

| Language | | | # of links | | |
|---|---|---|---|---|---|
| *Native* | *Foreign* | *Multiple* | *Internal* | *External* | *Total* |
| 8 | 5 | 14 | Min=0 | Min=0 | Min=0 |
| | | | Max=303 | Max=27 | Max=307 |
| | | | Total=2369 | Total=178 | Total=2547 |

## 5. Discussions

In this study, we examined cultural markers in a selected sample of Arabic web sites. The analysis revealed patterns of usage of cultural markers and in this section we describe how these findings relate to Hofstede's dimensions. Hofstede dimension says Arab culture has high power distance (80). This claim is supported by our findings which include the images of leaders (63%), images of official buildings (33%), and official logos (100%).Arab countries scoreless (38) in individualism and collectivism according to Hofstede dimension. Our findings which include the group pictures, lesser authentication passwords, supports Hofstede's claim. We found most websites have the pictures of males. Few of them also have the pictures of both males and females together, but females are covered in Abaya (traditional veil).This partially favors the Hofsetde's score for Arab countries, which says Arab countries have a score of 52 in masculinity versus femininity. The presence of simple menus, and detailed information supports Hofstede's claim that Arab countries have high (68) uncertainty avoidance. Hofstede dimension has no score for the long term versus short term orientation for Arab countries. We found 96% of the total websites surveyed have a search engine and 48% have site maps. Alumni links where found in fifty percent of the university websites. Moderate support for long term orientation was found based on site maps and alumni links. Our findings suggest short term orientation, however further research examining this factor/dimension is needed.

## 6. Conclusion

This study examined cultural markers of Arabic websites. Findings indicate that there are cultural markers which are largely influenced by the culture and the Hofstede's score for Arab countries is partially supported by the website design components examined in this study. Hofstede has no score for long term versus short term orientation, for which our analysis resulted in moderate support for long term orientation.

### Acknowledgment

This research project was supported by a grant from the Research Center of the Female colleges for Medical and Scientific studies in King Saud University. We are immensely thankful to Dr. Areej Al Wabil, without whose help this work would have been impossible.